# Coil Reweighting to Suppress Motion Artifacts in Real-Time Exercise Cine Imaging

Chong Chen, Yingmin Liu, Yu Ding, Matthew Tong, Preethi Chandrasekaran, Christopher Crabtree, Syed M. Arshad, Yuchi Han, Rizwan Ahmad*

The Ohio State University, Columbus OH, USA

*Corresponding author: ahmad.46@osu.edu

## Abstract

Background: Accelerated real-time cine (RT-Cine) imaging enables cardiac function assessment without the need for breath-holding. However, when performed during in-magnet exercise, RT-Cine images may exhibit significant motion artifacts.

Methods: By projecting the time-averaged images to the subspace spanned by the coil sensitivity maps, we propose a coil reweighting (CR) method to automatically suppress a subset of receive coils that introduces a high level of artifacts in the reconstructed image. RT-Cine data collected at rest and during exercise from ten healthy volunteers and six patients were utilized to assess the performance of the proposed method. One short-axis and one two-chamber RT-Cine series reconstructed with and without CR from each subject were visually scored by two cardiologists in terms of the level of artifacts on a scale of 1 (worst) to 5 (best).

Results: For healthy volunteers, applying CR to RT-Cine images collected at rest did not significantly change the image quality score (p=1). In contrast, for RT-Cine images collected during exercise, CR significantly improved the score from 3.9 to 4.68 (p<0.001). Similarly, in patients, CR did not significantly change the score for images collected at rest (p=0.031) but markedly improved the score from 3.15 to 4.42 (p<0.001) for images taken during exercise. Despite lower image quality scores in the patient cohort compared to healthy subjects, likely due to larger body habitus and the difficulty of limiting body motion during exercise, CR effectively suppressed motion artifacts, with all image series from the patient cohort receiving a score of four or higher.

Conclusion: Using data from healthy subjects and patients, we demonstrate that the motion artifacts in the reconstructed RT-Cine images can be effectively suppressed significantly with the proposed CR method.

## List of Abbreviations

CR, coil reweighting; RT, real-time; CMR, cardiovascular magnetic resonance imaging; ExCMR, exercise stress magnetic resonance imaging; ESPIRiT, an eigenvalue approach to autocalibrating parallel MRI; SCoRe, sparsity adaptive reconstruction; SAX, short-axis; 2CH, two-chamber; TE, echo-time; TR, repletion time; GRO, golden ratio offset

**Introduction**

Exercise stress cardiovascular MRI (ExCMR) provides valuable insights into cardiac function under stress conditions, offering a more comprehensive evaluation compared to resting conditions [1]. Since breath-holding is generally not feasible when imaging during exercise, accelerated real-time cine (RT-Cine) imaging is often employed to collect data under free-breathing conditions [2,3]. However, this approach is prone to significant motion artifacts, especially during in-magnet exercise [4], where extensive movement of body coils can degrade image quality due to changing coil sensitivity maps over time.

The artifacts caused by the physical movement of coils are typically visible in the time-averaged coil images. To suppress the motion artifacts, the data from specific coils could be excluded by visual inspection of the individual coil images. However, visual inspection is not practical in routine clinical use. In this work, we introduce a coil reweighting (CR) method to automatically minimize the impact of coils that contribute significant artifacts to the reconstructed image. Employing RT-Cine data collected both at rest and during in-magnet exercise from ten healthy volunteers and six patients, we demonstrate that the proposed CR method effectively reduces motion artifacts in the reconstructed RT-Cine images.

**Methods**

MRI signal acquired using a receive coil array resides in the subspace spanned by the coil sensitivity maps. Some torso movement is unavoidable during in-magnetic exercise; this movement, along with exaggerated breathing motion, may result in the physical movement of some of the receive coils. In principle, one could estimate temporally varying coil sensitivity maps using a sliding window approach. However, computing a large number of sensitivity maps slows down image reconstruction. Additionally, due to the high acceleration rates used in ExCMR, it is generally not feasible to keep the length of the sliding window short enough to "freeze" rapid body movements. Consequently, the sensitivity maps for ExCMR are typically estimated from the time-averaged k-space. While this approximation is reasonable when there is little to no torso movement, it can result in significant motion artifacts during ExCMR when both respiratory and bulk motions are pronounced.

To address this issue, we propose the pipeline shown in Figure 1(a) to automatically suppress the contribution from coils that introduce most of the artifacts. We first average the k-space of all the frames to generate the time-averaged k-space for each of the $N$ coils. The coil sensitivity maps $\{S_i\}_{i=1}^{N}$ are then estimated using ESPIRiT [5], and the time-averaged coil images $\{x_i\}_{i=1}^{N}$ are derived by performing inverse Fourier transformation of the time-averaged k-space. Subsequently, the coil images are projected onto the subspace spanned by the coil sensitivity maps, i.e.,

$$\tilde{x}_i = \left(\sum_{n=1}^{N} x_n \odot S_n^*\right) \odot S_i,$$

where $\{\tilde{x}_i\}_{i=1}^{N}$ are the projected images from $N$ coils, "$\odot$" represents element-wise multiplication, and the superscript "*" represents element-wise conjugation. For an RT-Cine image series collected during exercise, Figure 1(b) shows the time-averaged and projected images from two physical coils, one with significant motion artifacts (Coil-4) and the other without (Coil-11). The difference between $x_i$ and $\tilde{x}_i$ can be mainly attributed to motion artifacts, as one should expect $x_i \approx \tilde{x}_i$ when there is no coil movement. We propose using the residual signal, $\|\tilde{x}_i - x_i\|_2$, to identify the coils that strongly contribute to the image artifacts. Figure 1(c) shows $\|\tilde{x}_i - x_i\|_2$ for all physical coils.

After computing $\|\tilde{x}_i - x_i\|_2$ for each coil, we select $M$ coils with the largest $\|\tilde{x}_i - x_i\|_2$ values. Coil weights for those $M$ coils are set as $w_i = C^2/\|\tilde{x}_i - x_i\|_2^2$, where $C$ represents the average residual signal of the $K$ coils with smallest residuals. The weights for the remaining $N - M$ coils are left unchanged. After applying the reweighting, we compress the k-space data to 12 virtual coils for faster processing. Then, the coil sensitivity maps are re-estimated using ESPIRiT [5], and the reconstruction is performed using a parameter-free SENSE-based CS method called SCoRe [3,6]. For comparison, the data without CR are also reconstructed using the same method. In this work, we chose $M = 10$ and $K = 15$.

The RT-Cine images, reconstructed with and without CR, were visually scored by two cardiologists for the level of image artifacts. The images were evaluated using a five-point scale: 1--Unusable, 2--Severe artifacts obscuring useful information, 3--Moderate artifacts with some loss of information but still diagnostic, 4--Minor artifacts with minimal loss of information, 5--No artifacts.

**Experiments**

Ten healthy volunteers (4 females, 32±9 years) were scanned under free-breathing conditions using a prototype balanced steady-state free-precession RT-Cine sequence on a 3T scanner (Vida, Siemens Healthcare, Erlangen, Germany). The volunteers were instructed to exercise on a supine cycle ergometer (Lode BV, Netherlands) with an initial resistance of 20 W. Resistance was increased in 20 W increments up to a maximum of 60 W [7]. Two volunteers were not able to complete the three-stage protocol and stopped after reaching 40 W. At rest and each exercise stage, a SAX stack covering the whole heart and a 2CH slice were acquired using an 18-channel body array coil combined with a 12-channel posterior spine array coil, resulting in $N = 30$ physical coils. The other imaging parameters are TE/TR 1.29/2.91—1.38/3.08 ms, flip angle 38-44 degrees, slice thickness 6 mm, spatial resolution 1.67x1.67—2.29x2.29 mm$^2$, temporal resolution 34.8-40.7 ms, acquisition time 6 s/slice, and acceleration rate 8-9 with the variable density Cartesian GRO sampling pattern [8].

Besides the healthy subjects, six patients (3 females, 60±13 years) were also imaged but with a slightly different exercise protocol. In patients, a 10 W workload increment was employed when the 20 W increase was considered excessively challenging by the patients. The maximum exercise intensities achieved by the patients were 60 W, 30 W, 40 W, 20 W, 30 W, and 20 W, respectively. The RT Cine acquisition for patients mirrored that of the healthy volunteers.

This study was approved by the Institutional Review Board, and written informed consent was obtained from all participants. RT-Cine data (one mid-ventricular SAX and one 2CH slice) collected at rest and during the last stage of the exercise were utilized to assess the performance of the proposed method.

## Results

Figure 2 illustrates representative 2CH and SAX images collected during the in-magnet exercise, reconstructed with CR (CR+) and without CR (CR-). The top row depicts images from a healthy volunteer, while the bottom row displays images from a patient. Despite some signal loss, especially on the chest wall, a significant reduction in motion artifacts is evident with CR, as indicated by the red arrows. The temporal profiles across the heart are also displayed, highlighting the suppression of artifacts inside the blood pool and the myocardium.

Visual scoring by two cardiologists for ten volunteers is summarized in Table 1. For imaging at rest, CR had no significant impact on the image quality and resulted in the same average score of 4.93. In contrast, during exercise imaging, the average score increased from 3.9 to 4.68 with CR. The improvement offered by CR was statistically significant in the case of exercise (p<0.001) and not at rest (p=1).

Visual scoring for six patients is summarized in Table 2. For imaging at rest, CR had no significant impact on the image quality, with average scores of 4.63 vs. 4.88. However, the average score increased from 3.15 to 4.42 with CR for the images collected during exercise. The improvement was statistically significant in the case of exercise (p<0.001) and not at rest (p=0.031, $\alpha = 0.05$). More importantly, all slices in the patient study received a score of four or higher after CR. Figure 3 shows the 2CH view from Patient #5 at rest, where significant motion artifacts were observed due to the patient's bulk motion during acquisition. CR effectively mitigated these artifacts, leading to a score increase from 3 to 4.5. This example highlights the potential of CR for suppressing motion artifacts outside of exercise imaging.

## Discussion

When RT-Cine imaging is performed during in-magnet exercise, the coil sensitivity maps vary with time due to significant respiratory and bulk motion. Initially, we attempted to utilize temporally varying coil sensitivity maps estimated from a sliding window of a small number of frames. However, this approach proved unsuccessful. A smaller window failed to provide a fully sampled region of sufficient size or quality for reliable estimation of coil sensitivity maps, while a larger window introduced artifacts similar to those from time-averaged sensitivity maps. To address these issues, we calculated the residual signal for each physical coil based on how well its measurements are confined to the subspace spanned by coil sensitivity maps. By utilizing the residual signal values, the proposed CR method effectively suppressed the contributions of artifact-inducing coils. Our preliminary findings show that CR offered marked improvement for RT-Cine performed during exercise while having no adverse effect on the images collected during rest. Since CR serves as a computationally efficient preprocessing step, it can be seamlessly integrated with any reconstruction method. We used the values of $M = 10$ and $K = 15$ for all datasets included in this study, but these parameters can be further optimized.

## Conclusion

We demonstrated that the motion artifacts in the reconstructed RT-Cine images were suppressed significantly with the proposed coil reweighting method in both healthy volunteers and patients imaged under exercise stress.

## Acknowledgment

This work was funded by R01-HL151697, R01-HL148103, and R01-HL135489.

## References

[1] Craven TP, Tsao CW, La Gerche A, Simonetti OP, Greenwood JP. Exercise cardiovascular magnetic resonance: development, current utility and future applications. Journal of Cardiovascular Magnetic Resonance. 2020;22(1):65. doi:10.1186/s12968-020-00652-w

[2] Liang D, Liu B, Wang J, Ying L. Accelerating SENSE using compressed sensing. Magnetic Resonance in Medicine. 2009;62(6):1574-1584. doi:10.1002/mrm.22161

[3] Chen C, Liu Y, Schniter P, et al. Sparsity adaptive reconstruction for highly accelerated cardiac MRI. Magnetic Resonance in Medicine. 2019;81(6):3875-3887. doi:10.1002/mrm.27671

[4] He B, Chen Y, Wang L, Yang Y, Xia C, Zheng J, Gao F. Compact MR-compatible ergometer and its application in cardiac MR under exercise stress: A preliminary study. Magnetic Resonance in Medicine. 2022;88(4):1927-1936.

[5] Uecker M, Lai P, Murphy MJ, et al. ESPIRiT — An Eigenvalue Approach to Autocalibrating Parallel MRI: Where SENSE meets GRAPPA. Magnetic Resonance in Medicine. 2014;71(3):990-1001. doi:10.1002/mrm.24751

[6] Ahmad R, Schniter P. Iteratively Reweighted \ell_1 Approaches to Sparse Composite Regularization. IEEE Transactions on Computational Imaging. 2015;1(4):220-235. doi:10.1109/TCI.2015.2485078

[7] Chandrasekaran P, Chen C, Liu Y, et al. Accelerated Real-time Cine and Flow under In-magnet Staged Exercise. Published online February 27, 2024. doi:10.48550/arXiv.2402.17877

[8] Joshi M, Pruitt A, Chen C, Liu Y, Ahmad R. Technical Report (v1.0)--Pseudo-random Cartesian Sampling for Dynamic MRI. Published online June 7, 2022. doi:10.48550/arXiv.2206.03630

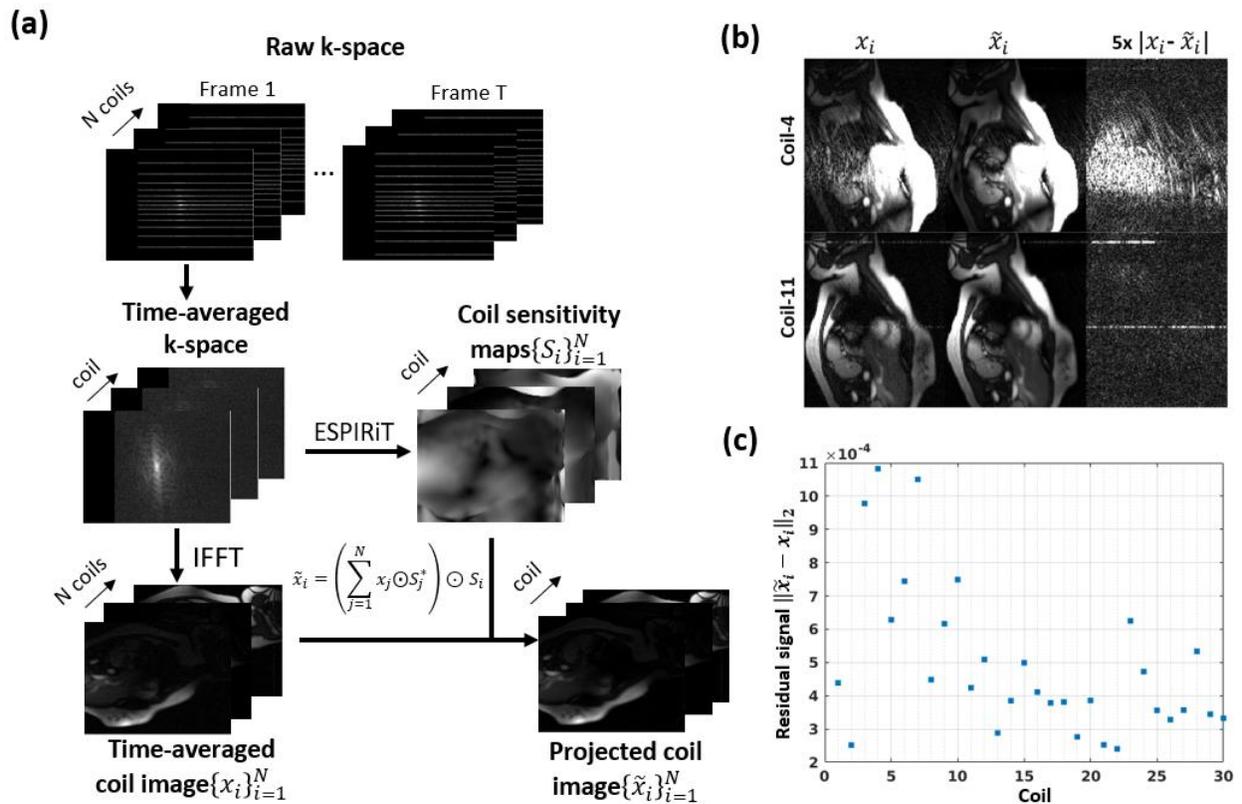

**Figure 1.** (a) Pipeline to generate time-averaged coil images $\{x_i\}_{i=1}^{N}$ and their projections $\{\tilde{x}_i\}_{i=1}^{N}$ to the subspace spanned by the ESPIRiT coil sensitivity maps $\{S_i\}_{i=1}^{N}$. (b). $x_i$, $\tilde{x}_i$ and their difference from two coils, one with significant motion artifacts (Coil-4) and one without (Coil-11), are shown. The horizontal lines in Coil-11 are likely from a faint RF interference. (c) The residual signal $\|\tilde{x}_i - x_i\|_2$ which cannot be characterized using ESPIRiT coil sensitivity maps.

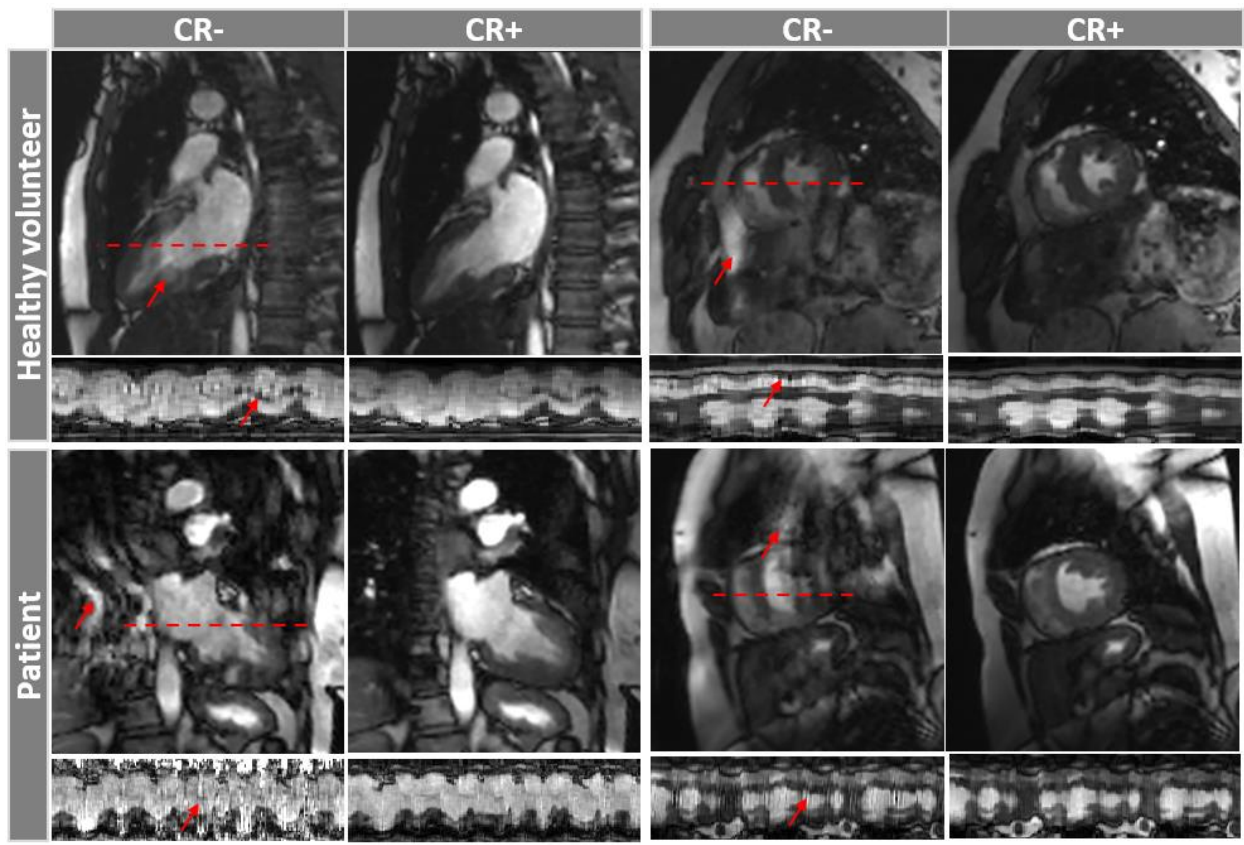

**Figure 2.** Exercise images reconstructed without (CR-) and with (CR+) coil reweighting. Temporal profiles along the red dashed lines are shown, with red arrows highlighting the motion artifacts. The top row displays images from a healthy volunteer, and the bottom row shows images from a patient.

**Table 1.** Artifact scores from healthy volunteers imaged at rest and during the last stage of exercise, showing average scores from two readers. Superior image quality is highlighted in bold.

| Volunteer | | | #1 | #2 | #3 | #4 | #5 | #6 | #7 | #8 | #9 | #10 | Average | |
|---|---|---|---|---|---|---|---|---|---|---|---|---|---|---|
| Rest (n=20) | CR- | 2CH | 5 | 5 | 5 | 5 | 5 | 5 | 4.5 | 5 | 5 | 4.5 | **4.9** | 4.93 |
| | | SAX | 5 | 4.5 | 5 | 5 | 5 | 5 | 5 | 5 | 5 | 5 | 4.95 | |
| | CR+ | 2CH | 4.5 | 5 | 5 | 5 | 5 | 5 | 4.5 | 5 | 5 | 4.5 | 4.85 | 4.93 |
| | | SAX | 5 | **5** | 5 | 5 | 5 | 5 | 5 | 5 | 5 | 5 | **5** | |
| Stress (n=20) | CR- | 2CH | 3.5 | 3.5 | 3.5 | 3.5 | 4 | 3.5 | 3.5 | 4 | 3.5 | 3.5 | 3.6 | 3.9 |
| | | SAX | 4.5 | 3.5 | 3.5 | 5 | 4.5 | 4 | 4 | **4.5** | 4.5 | 4 | 4.2 | |
| | CR+ | 2CH | **5** | **4.5** | **4.5** | **4.5** | **5** | **4.5** | **4.5** | 4 | **5** | **5** | **4.65** | **4.68** |
| | | SAX | **5** | **4** | **5** | **5** | **5** | **5** | **4.5** | 4 | **5** | **4.5** | **4.7** | |

**Table 2.** Artifact scores from patients imaged at rest and during the last stage of exercise, showing average scores from two readers. Superior image quality is highlighted in bold.

| Patient | | | #1 | #2 | #3 | #4 | #5 | #6 | Average | |
|---|---|---|---|---|---|---|---|---|---|---|
| Rest (n=12) | CR- | 2CH | 4.5 | 4.5 | 4.5 | 4.5 | 3 | 5 | 4.33 | 4.63 |
| | | SAX | 5 | 5 | 5 | 4.5 | 5 | 5 | 4.92 | |
| | CR+ | 2CH | **5** | 4.5 | 4.5 | **5** | 4.5 | 5 | **4.75** | 4.88 |
| | | SAX | 5 | 5 | 5 | **5** | 5 | 5 | **5.00** | |
| Stress (n=12) | CR- | 2CH | 2 | 3.5 | 2.5 | 3 | 3.5 | 3.25 | 2.96 | 3.15 |
| | | SAX | 3 | 3.5 | 2.5 | 4 | 3.5 | 3.5 | 3.33 | |
| | CR+ | 2CH | **4.5** | **4** | **4** | **4** | **5** | **4** | **4.25** | **4.42** |
| | | SAX | **4.5** | **4.5** | **4.5** | **4.5** | **5** | **4.5** | **4.58** | |

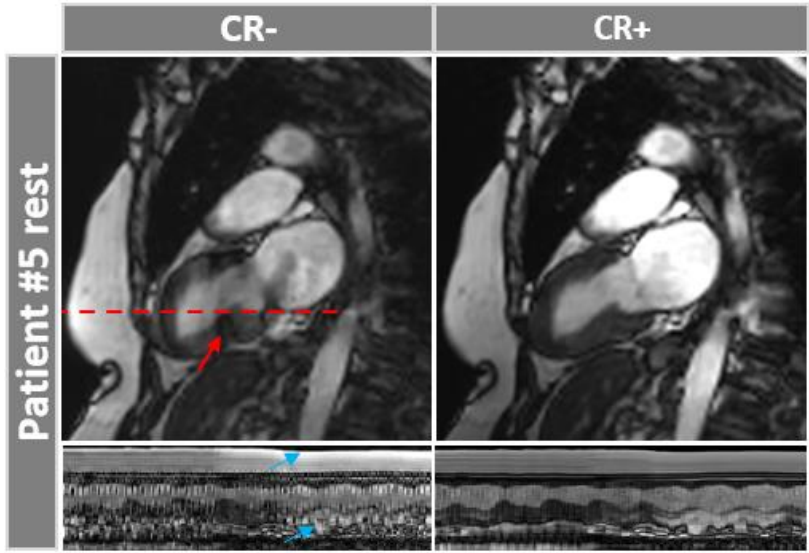

**Figure 3.** Rest images reconstructed without (CR-) and with (CR+) coil reweighting for Patient #5. Temporal profiles along the red dashed line are shown, with red arrows highlighting the motion artifacts. Blue arrows highlight the episode of bulk motion during the acquisition.